\documentclass{SCGE}
\usepackage[colorlinks,linkcolor=blue,citecolor=blue,urlcolor=blue]{hyperref}

\let\citedash\relax
\makeatletter \providecommand{\citedash}{\hbox{-}\penalty\@m}
\makeatother

\begin{document}

\begin{picture}(0,0){\rm
\put(0,-20){\makebox[160truemm][l]{\bf {\sanhao\raisebox{2pt}{.}}
Article  {\sanhao\raisebox{1.5pt}{.}}}}}
\put(0,-34){\jiuwuhao {\textcolor[rgb]{0.5,0.5,0.5}{\sf 
}}}
\end{picture}

\def\bm{\boldsymbol}
\def\dl{\displaystyle}
\def\du{\end{document}}
\def\d{{\rm d}}
\def\e{{\rm e}}
\def\i{{\rm i}}

\Year{2017} %
\Month{February} %
\Vol{60} 
\No{2} 
\BeginPage{1} 
\AuthorMark{{\rm H. Lin}, et al.}  
\DOI{10.1007/s11433-016-0464-3} 
\ArtNo{027411}

\title{Superconductivity in LiOHFeS single crystals with a shrunk c-axis lattice constant}

\author[]{Hai Lin}{}
\footnote{*Corresponding author (email: hhwen@nju.edu.cn)}
\author[]{Ruizhe Kang}{}
\author[]{Lu Kong}{}
\author[]{Xiyu Zhu}{}
\author[*]{Hai-Hu Wen}{}

\address[]{Center for Superconducting Physics and Materials,}
\address[]{National Laboratory of Solid State Microstructures and Department of Physics,}
\address[]{National Collaborative Innovation Center of Advanced Microstructures, Nanjing University, Nanjing 210093, China}

\maketitle \vspace{-3.5mm}{\footnotesize\begin{center} Received November 20, 2016; accepted November 25, 2016; published online December 2, 2016
\end{center}}\vspace*{-5mm}

\begin{center}
\rule{16.5cm}{0.4pt}
\parbox{16.5cm}
{\begin{abstract}
By using a hydrothermal ion-exchange method, we have successfully grown superconducting crystals of LiOHFeS with ${T_c}$ of about 2.8 K. Being different from the sister sample (Li${_{1-x}}$Fe${_x}$)OHFeSe, the energy dispersion spectrum analysis on LiOHFeS shows that the Fe/S ratio is very close to 1:1, suggesting an almost charge neutrality and less electron doping in the FeS planes of the system. Comparing with the non superconducting LiOHFeS crystal, each peak of the X-ray diffraction pattern of the superconducting crystal splits into two, and the diffraction peaks locating at lower reflection angles are consistent with that of non-superconducting ones. The rest set of diffraction peaks with higher reflection angles is corresponding to the superconducting phase, suggesting that the superconducting phase may has a shrunk c-axis lattice constant. Magnetization measurements indicate that the magnetic shielding due to superconductivity can be quite high under a weak magnetic field. The resistivity measurements under various magnetic fields show that the upper critical field is quite low, which is similar to the tetragonal FeS superconductor.
\end{abstract}}
\end{center}\vspace*{-0.6cm}

\begin{center}
\parbox{16.5cm}
{\bf\jiuhao iron chalcogenides, magnetic properties, transport properties}
\end{center}

\begin{center}
{}
\CITA    
\end{center}

\textwidth=178truemm \textheight=236truemm

\wuhao\vspace*{1.5mm}

\begin{multicols}{2}

\renewcommand{\baselinestretch}{1.08} \baselineskip 12.2pt\parindent=10.8pt

\renewcommand{\thefootnote}

\section{Introduction}\label{sec:3}

Since the high temperature superconductor LaFeAsO${_{1-x}}$F${_x}$\cite{HosonoJACS2008} was discovered in 2008, many different structures of iron-based superconductors have been found in succession\cite{ChuCW,WenHH,RGreene}. Therein, FeAs-based as well as FeSe-based superconductors are two most common and important families, which are named by their superconducting FeAs-layers or FeSe-layers. For iron-based superconductors, the highest superconducting transition temperature defined by Meissner effect is about 65K in monolayer FeSe film grown on SrTiO$_3$ substrate\cite{Xueqikun}. This has drawn great attention and leaded to numerous studies on FeSe-based superconductors with different layer structures. Concerning the charge neutrality and weak Van der Waals force of FeSe layers, more FeSe-based superconductors are created by intercalation, these include $A$${_x}$Fe${_{2-y}}$Se${_2}$ ($A$ is alkali or alkali-earth metal)\cite{XiaolongChenKFeSe}, (Li${_{1-x}}$Fe${_x}$)OHFeSe\cite{ChenxianhuiLiOHFeSe}, $A_x$(NH$_3$)$_y$Fe$_{2-z}$Se$_2$\cite{ammoniaKFeSe}, etc.. On the other hand, since sulfur and selenium both have the "2-" valence state, people were trying to use sulfur to substitute selenium in FeSe-layers. For example, slight sulfur doping in Fe${_{1+x}}$Se can raise the superconducting transition temperature ${T_c}$\cite{FeSeS}, and tetragonal FeS was also found to be a superconductor with $T_{\mathrm{c}}$ about 4.5 K\cite{HuangfuqiangFeS}. The crystal structure and band structure of tetragonal FeS are quite similar to FeSe\cite{FongChiHsuFeSe,FeSeband}. However, according to literature, with increasing the doping level of sulfur on (Li${_{0.8}}$Fe${_{0.2}}$)OHFeSe, $T_{\mathrm{c}}$ decreases monotonically and superconductivity is totally killed in LiOHFeS\cite{ChenxianhuiLiOHFeSeS,HuangfuqiangLiOHFeS,JohrendtLiOHFeS}. Therefore, though LiOHFeS has the same crystal structure as (Li${_{1-x}}$Fe${_x}$)OHFeSe, no superconductivity has been found in it so far. Is there any chance to get superconductivity in LiOHFeS? Furthermore, what is the dominant factor of superconductivity? These questions are very interesting and worth exploring.

In this work, we report the independent discovery of superconductivity in the crystals of LiOHFeS synthesized by hydrothermal method. The samples all show two sets of X-ray diffraction peaks. And each set of peaks represents a self-consistent c-axis lattice constant. Comparing with non superconducting samples, the superconducting ones have a new set of peaks with higher reflection angles, which might be responsible for superconductivity.
We have measured magnetization and electrical resistivity of the synthesized LiOHFeS crystal, and also present the relation between crystal structure and superconductivity. However, we must mention that, when preparing the paper, we notice that another group also shows superconductivity in LiOHFeS very recently\cite{arxivLiOHFeS} although the methods for fabricating the superconducting samples are very different.

\section{Experimental details}\label{sec:4}
In this work, both superconducting(SC) and non-superconducting(non-SC) LiOHFeS crystals
were synthesized by using a hydrothermal ion-exchange method, which is similar to (Li${_{1-x}}$Fe${_x}$)OHFeSe and FeS single crystals as reported previously\cite{Dong,LinhaiLiOHFeSe,LinhaiFeS}. First the precursor K${_{0.8}}$Fe${_{1.6}}$S${_2}$ single crystals are grown by the self-flux method. For the non-SC LiOHFeS crystals, we put 5g LiOH (J${\&}$K, 99${\%}$ purity), 0.6 g iron powder (Aladdin Industrial, 99.99${\%}$ purity), 0.2 g thiourea (J${\&}$K, 99.9${\%}$ purity) and 40 mg K${_{0.8}}$Fe${_{1.6}}$S${_2}$ single crystals into 10 mL deionized water in a teflon-lined stainless-steel autoclave (volume 50 mL). The autoclave was sealed and heated up to 120$^{\circ}$C followed by staying for 25 hours. In contrast, we made the SC LiOHFeS crystals in two steps. Firstly. we put 3 g NaOH, 0.6 g iron powder, 0.2 g thiourea and $\sim$ 1 g K${_{0.8}}$Fe${_{1.6}}$S${_2}$ single crystals into 10 mL deionized water and heated at 100 $^{\circ}$C for 25 hours. Then, after cooling down to room temperature, we put extra 5 g LiOH and 5 mL water into this autoclave and heated it up again to 120 $^{\circ}$C, staying for another 24 hours. The samples in the reacting solution were obtained after leaching. They have metallic lustrous surfaces.

The X-ray diffraction (XRD) measurements were performed on a Bruker D8 Advanced diffractometer with the Cu-K$_\alpha$ radiation. DC magnetization measurements were carried out with a quantum design instrument SQUID-VSM-7T. The resistive measurements were done with the standard four-probe method on a Quantum Design
instrument Physical Property Measurement System (PPMS). Scanning electron microscope (SEM) pictures and energy dispersive X-ray spectrum (EDS) measurements were performed at an accelerating voltage of 20kV and working distance of 10 millimeters by a scanning electron microscope (Hitachi Co.,Ltd.). In this paper, all measurements for SC LiOHFeS crystals were performed on the samples from the same batch.

\section{Results and discussion}\label{sec:5}
\begin{figure}[H]
\centering
\includegraphics[width=8.5cm]{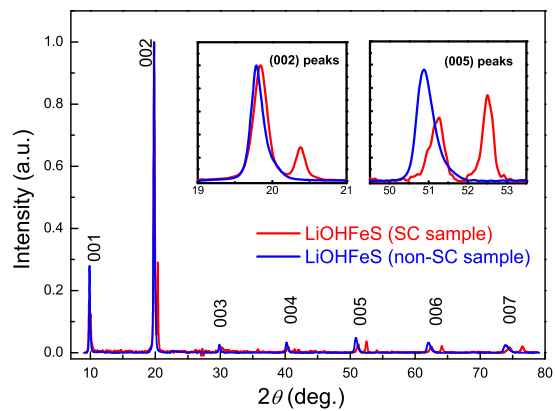}
\caption{(Color online) X-ray diffraction patterns for the SC(red lines) and non-SC(blue lines) LiOHFeS samples. One can see the predominant
(00$l$) indices and high coincidence of these two samples. The insets show two enlarged views near the (002) and (005) peaks, showing an obvious split of XRD reflection peaks for the SC LiOHFeS sample.} \label{fig1}
\end{figure}

Fig.~\ref{fig1} shows the XRD spectra of SC and non-SC LiOHFeS single crystals. For the non-SC single crystal, only ($00l$) reflections can be observed, indicating highly orientation along the \emph{c}-axis. The c-axis lattice constant determined here is about 8.96(4)$\AA$, close to the previously reported non-SC samples\cite{HuangfuqiangLiOHFeS}. However, for the SC single crystal, all the XRD reflection peaks split into two sets of diffraction peaks. And this split can be seen more clearly with enlarged views of 002 and 005 peaks, as shown in the insets of Fig.~\ref{fig1}. Each set of peaks can be well indexed with one group of ($00l$) reflections. Hence, two different c-axis lattice constants are obtained on the SC samples. The larger $c$ = 8.91(4) $\AA$ is almost equal to that of the non-SC sample, while another c-axis lattice constant with 8.71(1) $\AA$ is much smaller than that of the non-SC sample.
Consequently, it is reasonable to conclude that the lattice with larger $c$-axis lattice constant represents the non-SC part, while the lattice with smaller $c$-axis lattice constant is responsible for superconductivity. And the difference between SC and non-SC samples is mainly induced by different ratios of K${_{0.8}}$Fe${_{1.6}}$S${_2}$ crystals and LiOH, together with different growing environments. Notably, there are no diffraction peaks of tetragonal FeS crystals at all, which is verified by measuring many surfaces and inner surfaces of our samples.

\begin{figure}[H]
\centering
\includegraphics[width=8.5cm]{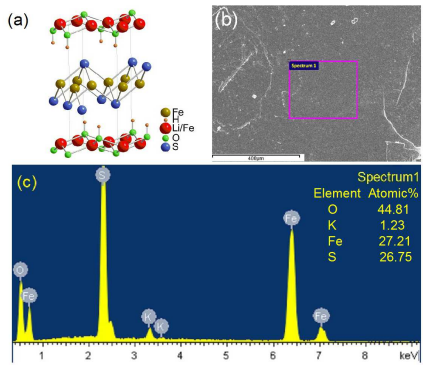}
\caption{(Color online) (a) Schematic structure of LiOHFeS. (b) SEM photograph of a SC LiOHFeS sample. (c) Energy dispersive X-ray microanalysis spectrum
taken on the SC LiOHFeS crystal shown in (b). The inset shows the atomic percentage of various elements.
} \label{fig2}
\end{figure}

In Fig.~\ref{fig2}(a), we show the schematic structure of LiOHFeS, which is similar to that of previously reported (Li${_{1-x}}$Fe${_x}$)OHFeSe. As shown in Fig.~\ref{fig2}(b), the SEM image from a SC LiOHFeS sample displays a very flat surface. This reveals the good quality of these LiOHFeS crystals. Then we performed an EDS measurement inside the red rectangle and obtained the atomic ratios, as shown in Fig.~\ref{fig2}(c). Lithium and hydrogen atoms can not be detected by EDS, so we cannot tell the exact content of lithium and hydrogen. We also want to mention that the concentration of oxygen shown here does not reflect the real value of oxygen in the sample because of the possible contamination from the substrat. As we can see, there are tiny amount of potassium left, which may come from the residual unreacted K${_{0.8}}$Fe${_{1.6}}$S${_2}$. This K${_{1-x}}$Fe${_{2-y}}$S${_2}$ has never been found to be superconductive. And it is notable that the sodium content is always zero in all these samples, ruling out the possible effect due to (Na${_{1-x}}$Fe${_x}$)(OH)$_2$FeSe\cite{arxivLiOHFeS}. The atomic percentages of iron and sulfur are nearly the same, with Fe:S = 1.017 : 1. We also measured the atomic ratios of other five samples from various batches, the Fe/S ratios are 0.952, 0.996, 1.017, 1.040, 1.047, respectively. It is clear that all of them are close to 1. So we believe our SC LiOHFeS crystals are nearly electric neutral, which suggests an ionic state of Fe$^{2+}$ in the FeS layer. This is different from its sister sample (Li${_{1-x}}$Fe${_x}$)OHFeSe, where the Fe/Se ratios normally remain in 1.2\cite{ChenxianhuiLiOHFeSe}. Since the extra 0.2Fe is doped to the Li sites in (Li${_{1-x}}$Fe${_x}$)OHFeSe, and the ratio of Fe:Se is about 0.98:1 on the FeSe layer\cite{Dong}, therefore we believe that the FeSe layer is more electron doped. The electric neutrality in LiOHFeS indicates that the Li sites are probably not substituted by Fe in our present sample, or the amount of iron in the Li sites is almost equal to the number of iron vacancies in FeS-layer. The possible fluctuation of the self-doping between Li and Fe, although very limited, in the LiOH block may be the cause for the different $c$ lattice constant and also superconductivity. But unfortunately since the doping of Fe to the Li site in our LiOHFeS sample is difficult, this may explain why our LiOHFeS crystal is superconducting but its $T_{\mathrm{c}}$ is lower than that in other work\cite{arxivLiOHFeS}.

\begin{figure}[H]
\centering
\includegraphics[width=8.5cm]{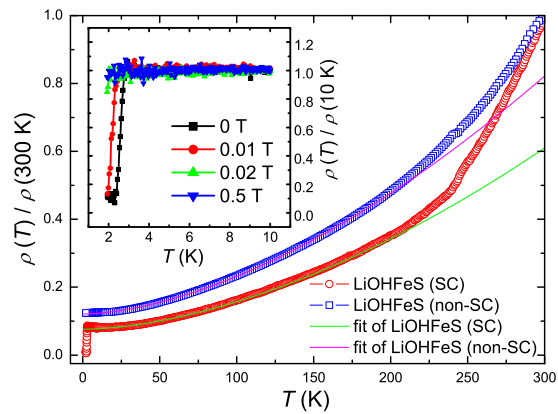}
\caption{(Color online) Temperature dependence of resistivity for the SC(red lines) and non-SC(blue lines) LiOHFeS crystals at zero field with measuring current of 100 $\upmu$A. The solid lines are the fits in the low temperature range by $\rho(0)$+\emph{AT}${^n}$. The upper left inset is the temperature of resistivity for the SC crystal at various fields parallel to c-axis. Superconductivity can be easily suppressed by a very low field.
} \label{fig3}
\end{figure}

We present the electrical resistivity measurements in Fig.~\ref{fig3}, showing the temperature dependence of resistivity for LiOHFeS crystals at zero field with measuring current of 100 $\upmu$A. The open circles are data of the SC LiOHFeS sample and open squares are data of the non-SC sample. Both resistivities decrease monotonically when cooling down to low temperature and show a highly metallic conductivity.
For non-SC LiOHFeS, this metallic-like behavior is different from the previous work of non-SC (Li${_{1-x}}$Fe${_x}$)OHFeS that behaves like a semiconductor at low temperatures\cite{HuangfuqiangLiOHFeS}. This suggests that even in the same system, a little difference in crystal structure or stoichiometry can obviously influence the transport properties. It indicates that LiOHFeS may be superconducting by simply changing synthesis condition, which has actually been demonstrated in this work.
For SC LiOHFeS, a sharp superconducting transition is observed at $T_{\mathrm{c}}^{\mathrm{onset}}$ = 2.8 K. Furthermore, the curvature of resistivity from low temperature to 300 K is extremely positive and the curvature seems larger and larger. This divergent behavior of normal state resistivity is similar to superconducting FeS system\cite{HuangfuqiangFeS}, however, different from the FeSe-layer systems. In its sister sample (Li${_{1-x}}$Fe${_x}$)OHFeSe, or Fe${_{1+x}}$Se single crystal, the positive curvature becomes smaller in high temperature region\cite{ChenxianhuiLiOHFeSe,FongChiHsuFeSe}. And in $K$${_x}$Fe${_{2-y}}$Se${_2}$ system, the curvature will even become negative at 105 K\cite{XiaolongChenKFeSe} and a dome-like feature even appears. This extreme positive curvature of $\rho$ vs. $T$ in wide temperature region is quite strange, which either suggests a very high Debye temperature or an unusual scattering mechanism. The solid lines are the fits to a power-law in low temperature region ranging from 10 K to 200 K by the formula $\rho(T)$=$\rho(0)$+\emph{A}$T$${^n}$, and one can see that the data of both SC and non-SC LiOHFeS crystals are fitted very well in a wide temperature range. The pink solid line fits for non-SC LiOHFeS data with \emph{n}= 1.67408. The green solid line fits the data for SC LiOHFeS with \emph{n} = 1.71589. And the residual resistivity ratio, defined as
RRR=$\rho$(300K)/$\rho$(0K) = 13, is quite large, indicating the low concentration of impurity or vacancies in the sample.
The inset of Fig.~\ref{fig3} shows the temperature dependence of resistivity at various magnetic fields from 1.9 K to 10 K. Resistivity of normal state has little change even under 0.5 T, revealing a nearly zero magnetoresistance. One can see $T_{\mathrm{c}}^{\mathrm{onset}}$ = 2.8 K is easily suppressed down to 2 K at 0.02 T. This extreme sensitivity to magnetic field indicates the upper critical field $H_c$ is very low, similar to the tetragonal FeS system\cite{HuangfuqiangFeS}.

\begin{figure}[H]
\centering
\includegraphics[width=8.5cm]{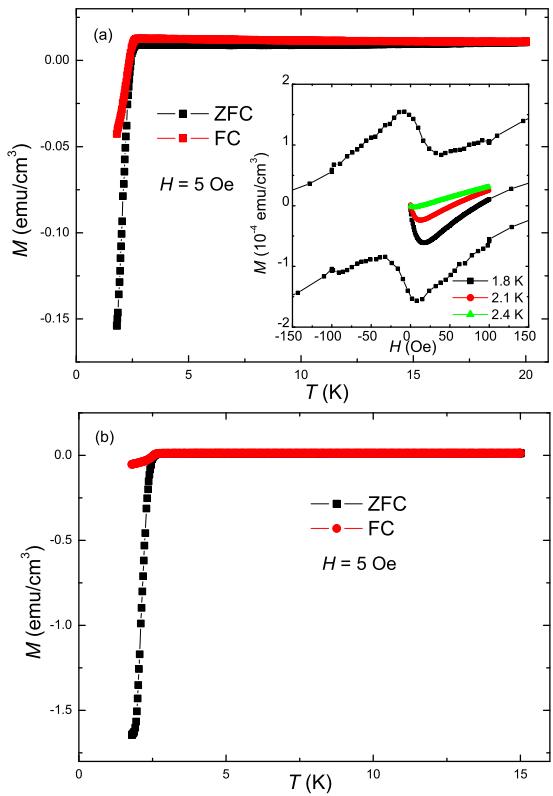}
\caption{(Color online) (a) Temperature dependence of magnetic susceptibility for the SC LiOHFeS crystal measured in both ZFC(black line) and FC(red line) modes, with an applied field of 5 Oe parallel to c-axis. The lower left inset shows the MHLs at 1.8 K, 2.1 K and 2.4 K, which are originally measured between $\pm$3000 Oe, but only the enlarged views between $\pm$150 Oe are shown here. (b) Temperature dependence of magnetic susceptibility for another SC LiOHFeS sample, which shows a larger diamagnetic signal. }
\label{fig4}
\end{figure}

In Fig.~\ref{fig4}(a), we demonstrate the magnetic susceptibility of the SC LiOHFeS crystal from 1.8 K to 20 K with applying an external field of 5 Oe parallel to c-axis. The SC LiOHFeS sample is measured in both the zero-field-cooled (ZFC) and field-cooled(FC) modes. A sharp superconducting transition is obvious at 2.7 K from the susceptibility measurements, which is corresponding well to the resistivity data. The maximum diamagnetic signal is about -0.154 emu/cm$^3$, or 4$\pi$$\chi$ = -0.39. Due to the low temperature limit of our instrument SQUID-VSM, we can not reach the complete diamagnetism. But for some other samples, the largest 4$\pi$$\chi$ of our samples can reach to -4.1, indicating a large superconducting volume in our SC LiOHFeS samples, as shown in Fig.~\ref{fig4}(b). The inset in Fig.~\ref{fig4}(a) shows the magnetization hysteresis loops(MHLs) of this SC LiOHFeS sample at various temperatures below $T_{\mathrm{c}}$. The MHLs are originally measured between $\pm$3000 Oe. For a better demonstration, we show the enlarged views of MHLs between $\pm$150 Oe here. The sweep rate of field below 100 Oe is 2 Oe/s and the one higher than 100 Oe is as fast as 10 Oe/s. Strong magnetic relaxation leads to the incoincidence of the MHLs in low field region. Taking a look at the magnetic field penetration process, one can see that the magnetic field intrudes into the superconductor very easily. This is similar to the data in tetragonal FeS system\cite{HuangfuqiangFeS,LinhaiFeS}. The "lower critical field" at about 15 Oe at 1.8 K seems too low, but this may be induced by the large demagnetization effect of the very flat shape of the sample when the field is perpendicular to the basal plane.

\section{Conclusions}\label{sec:5}
In summary, we successfully synthesized the superconducting LiOHFeS crystals with large size and good quality by the hydrothermal ion-exchange method. The Fe/S ratio of these LiOHFeS samples is very close to 1:1, as shown in the EDS analysis. It is different from (Li${_{1-x}}$Fe${_x}$)OHFeSe, suggesting very weak doping of Fe to the Li sites. The XRD patterns show an obvious split of all diffraction peaks in the SC samples, indicating two phases with different c-axis lattice constants. One set of diffraction peaks locating at lower reflection angles has a larger $c$ = 8.91 $\AA$, which is very close to non superconducting LiOHFeS with $c$ = 8.96 $\AA$. But the other set of peaks locating at higher reflection angles has a much smaller $c$ = 8.71 $\AA$, indicating another phase with a shrunk $c$-axis lattice constant. This phase is probably the source of superconductivity. Magnetization measurement displays a sharp superconducting transition at $T_{\mathrm{c}}$ = 2.8 K and a large magnetic screening volume. Resistivity under various magnetic fields reveals a quite low upper critical field, which is similar to the tetragonal FeS system.

\vspace*{2mm} \Acknowledgements{\bahao This work was supported by the National Natural Science Foundation of China (Grant Nos. A0402/11534005, and A0402/11190023), and the Ministry of Science and Technology of China (Grant Nos. 2016YFA0300401, 2016FYA0401704 , and 2012CB821403).}

\end{multicols}
\end{document}